\documentclass[fleqn,usenatbib]{mnras}

\usepackage{newtxtext,newtxmath}

\usepackage[T1]{fontenc}
\usepackage{ae,aecompl}

\usepackage{graphicx}	
\usepackage{amsmath}	
\usepackage{color}      

\newcommand{\ie}{{\it i.e.}}

\newcommand{\be}{\begin{equation}}
\newcommand{\ee}{\end{equation}}

\title[Galaxy Zoo: 3D]{Galaxy Zoo: 3D - Crowd-sourced Bar, Spiral and Foreground Star Masks for MaNGA Target Galaxies}

\author[Masters et al. ]{Karen L. Masters$^{1}$\thanks{E-mail: klmasters@haverford.edu},  Coleman Krawczyk$^2$, Shoaib Shamsi$^{1}$, Alexander Todd$^{2,3}$, \newauthor Daniel Finnegan$^{1,4}$, Matthew Bershady$^{5,6,7}$, Kevin Bundy$^{8}$, Brian Cherinka$^{9}$, \newauthor  Amelia Fraser-McKelvie$^{10,11}$, Dhanesh Krishnarao$^{9}$, Sandor Kruk$^{12}$, \newauthor Richard R. Lane$^{13}$,  David Law$^{9}$, Chris Lintott$^{14}$,   Michael Merrifield$^{15}$, \newauthor Brooke Simmons$^{16}$,  Anne-Marie Weijmans$^{17}$,  Renbin Yan$^{18}$
\\
$^{~1}$Departments of Physics and Astronomy, Haverford College, 370 Lancaster Avenue, Haverford, Pennsylvania 19041, USA\\
$^{~2}$Institute of Cosmology \& Gravitation, University of Portsmouth, Dennis Sciama Building, Portsmouth, PO1 3FX, UK\\
$^{~3}$University of Bath, Claverton Down, Bath, BA2 7AY, UK\\
$^{~4}$Keck Northeast Astronomy Consortium (KNAC) REU Student\\
$^{~5}$Department of Astronomy, University of Wisconsin-Madison, Madison, WI~53706, USA\\
$^{~6}$South African Astronomical Observatory, PO Box 9, Observatory 7935, Cape Town, South Africa\\
$^{~7}$Department of Astronomy, University of Cape Town, Private Bag X3, Rondebosch 7701, South Africa\\
$^{~8}$UCO/Lick Observatory, University of California, Santa Cruz, 1156 High St. Santa Cruz, CA 95064, USA\\
$^{~9}$Space Telescope Science Institute, 3700 San Martin Drive, Baltimore, MD 21218, USA\\
$^{10}$International Centre for Radio Astronomy Research, The University of Western Australia, 35 Stirling Hwy, 6009 Crawley, WA, Australia \\
$^{11}$ARC Centre of Excellence for All Sky Astrophysics in 3 Dimensions (ASTRO 3D) \\
$^{12}$European Space Agency, ESTEC, Keplerlaan 1, NL 2201 AZ, Noordwĳk, The Netherlands\\
$^{13}$Centro de Investigaci\'on en Astronom\'ia, Universidad Bernardo O'Higgins, Avenida Viel 1497, Santiago, Chile\\
$^{14}$Oxford Astrophysics, Department of Physics, University of Oxford, Denys Wilkinson Building, Keble Road, Oxford, OX1 3RH, UK\\
$^{15}$School of Physics \& Astronomy, University of Nottingham, University Park, Nottingham, NG7 2RD, UK\\
$^{16}$Department of Physics, Lancaster University, Bailrigg, Lancaster, LA1 4YB, UK\\
$^{17}$School of Physics and Astronomy, University of St Andrews, North Haugh, St Andrews KY16 9SS, UK\\
$^{18}$Department of Physics and Astronomy, University of Kentucky, 505 Rose St., Lexington, KY 40506-0057, USA}

\date{\today}

\pubyear{2020}

\hypersetup{draft}
\begin{document}
\label{firstpage}
\pagerange{\pageref{firstpage}--\pageref{lastpage}}
\maketitle

\begin{abstract}
The challenge of consistent identification of internal structure in galaxies -- in particular disc galaxy components like spiral arms, bars, and bulges -- has hindered our ability to study the physical impact of such structure across large samples.  In this paper we present {\it Galaxy Zoo: 3D} (GZ: 3D) a crowdsourcing project built on the {\it Zooniverse} platform which we used to create spatial pixel (spaxel) maps that identify galaxy centres, foreground stars, galactic bars and spiral arms for 29831 galaxies which were potential targets of the MaNGA survey (Mapping Nearby Galaxies at Apache Point Observatory, part of the fourth phase of the Sloan Digital Sky Surveys or SDSS-IV), including nearly all of the 10,010 galaxies ultimately observed. Our crowd-sourced visual identification of asymmetric, internal structures provides valuable insight on the evolutionary role of non-axisymmetric processes that is otherwise lost when MaNGA data cubes are azimuthally averaged. We present the publicly available GZ:3D catalog alongside validation tests and example use cases. These data may in the future provide a useful training set for automated identification of spiral arm features. As an illustration, we use the spiral masks in a sample of 825 galaxies to measure the enhancement of star formation spatially linked to spiral arms, which we measure to be a factor of three over the background disc, and how this enhancement increases with radius.

\end{abstract}

\begin{keywords}
galaxies: bar -- galaxies: spiral -- galaxies: structure -- surveys -- methods: data analysis
\end{keywords}

\section{Introduction}
We are in an era of large scale surveys of resolved spectroscopy, with a number of galaxy surveys aiming to obtain Integral Field Unit (IFU; sometimes Integral field spectrograph, IFS) data for large samples of galaxies. While the largest of these to date is Mapping Nearby Galaxies at Apache Point Observatory (MaNGA, \citealt{Bundy2015}; part of the Sloan Digital Sky Surveys, or SDSS-IV, \citealt{Blanton2017}), it follows a long history of similar surveys (e.g SAURON, \citealt{deZeeuw2002}; ATLAS-3D, \citealt{Cappellari2011}; CALIFA, \citealt{Sanchez2012}; DiskMass, \citealt{Bershady2010}; SAMI, \citealt{Bryant2015}). 

The promise of IFU spectroscopy is that, through producing spatially resolved spectral cubes, the observations can reveal the details of how galaxy evolution proceeds inside statistically significant samples of nearby galaxies. These data provide both stellar and gaseous kinematics and maps of emission line properties. These data can also be used, via stellar population modelling (e.g. Pipe3D; \citealt{sanchez2016b}) to create maps of chemical composition, stellar population ages, star-formation rates (SFR) and more. However visualisation of large samples of these highly dimensional data is a challenge. 

Many authors resort to parameterising the complex three-dimensional (3D) data into azimuthally averaged radial gradients. For example, several works all using MaNGA data, have looked at azimuthally averaged stellar population parameters \citep[e.g.][]{Li2015, IbarraMedel2016,goddard2017,Wang2018,Chen2019,Chen2020,Parikh2021}, or gas properties \citep[e.g.][]{Belfiore2017,Belfiore2018}. Many of these works \citep[e.g.][]{goddard2017,Wang2018,Chen2020,Parikh2021} do comment on a difference in gradients linked to galaxy morphology (e.g. early-types tended to have flatter gradients than late types), but in general they do not consider the systematics introduced into azimuthally averaged radial gradients by non-axisymmetric features such as bars, and spiral arms. 

IFU data, such as that provided by MaNGA have the potential to go beyond these two-dimensional views of galaxies. Nearby galaxies are not point sources, and most are not azimuthally symmetric or smooth. For example Galaxy Zoo \citep{Lintott2011} revealed that most galaxies in the SDSS Main Galaxy Sample showed spiral structure, and many of these disc galaxies contain bars - significant central non-axisymmetric structure, e.g. \citet{Masters2011bars} find 29.4\% of {\it Galaxy Zoo} spirals have strong bars. In addition, the presence of foreground stars and foreground or background galaxies can present a challenge for the automated data reduction pipelines needed for large samples. 

 It was considerations such as these which inspired the idea for the {\it Galaxy Zoo: 3D} (GZ:3D) project presented in this paper, with the goal of identifying internal structures, galaxy centres (whether the main target or additional galaxies), and foreground stars in MaNGA target objects. GZ:3D is a citizen science project that provides bar, spiral arm, central, and foreground star masks for almost all MaNGA targets galaxies. Used in conjunction with MaNGA galaxy survey data, it allows for the selection of various regions of interest from the IFU data cubes and associated analysis output maps. 
 
 {\it Galaxy Zoo: 3D} does not provide information to separate bulge and disc components of galaxies. Other techniques which use semi-automated 2D decomposition, based on images \citep{CatalanTorrecilla2017,Kruk2018}, or spectral data cubes themselves \citep{Tabor2019}, are able to do this. However multi-component decomposition including both bars and spirals is beyond the ability of most codes, without significant human intervention (see \citealt{Lingard2020} for an attempt to leverage Galaxy Zoo style crowd-sourcing to help with this challenge). Even neglecting a strong bar component can lead to a bias in the bulge-to-total ratios for barred galaxies (e.g. as shown by \citealt{MendezAbreu2017,Kruk2018}). The GZ:3D technique was developed to leverage human pattern recognition, and provide a guide to the location of complex internal structures in MaNGA data.

We introduce the MaNGA survey, SDSS images and the building of the {\it Galaxy Zoo: 3D} site on the {\it Zooniverse} Project Builder in Section \ref{sec:methods}. This section also contains details of how the crowd-sourced classfications are aggregated and turned into spatial pixel (spaxel) masks (for bars and spirals) or clustered positions (for galaxy centres and foreground stars). Section \ref{sec:results} gives an overview of the output from {\it Galaxy Zoo: 3D} with a subsection on each of the types of identified objects and structures. As well as showing some example data and uses, this section provides an overview of how {\it Galaxy Zoo: 3D} results have been used to date in published works. As an example usage of GZ:3D we present new results on the fraction of star formation (SF) found inside spiral arm masks in MaNGA galaxies, as well as revealing how this enhancement increases with radius. We provide a section on data access and some advice on how the reader might use the GZ:3D masks in combination with MaNGA (Section \ref{sec:data}) and finish with a short summary and conclusions (Section \ref{sec:summary}). 

Where physical units are employed we use a value of $H_0 = 70 $ km/s/Mpc. 

\section{Methods} \label{sec:methods}

\subsection{Sample Selection}
 The goal for the {\it Galaxy Zoo: 3D} project was to provide masks for use with all galaxies observed by the MaNGA survey. The MaNGA survey, part of SDSS-IV \citep{Blanton2017}, used the MaNGA IFU bundles \citep{Drory2015} and the BOSS Spectrograph \citep{Smee2013} on the Sloan Foundation 2.5m telescope \citep{Gunn2006} to observe nearby galaxies, with the target selection described in \citet{Wake2017}. For details of MaNGA observing, reduction and calibration see \citet{Law2015,Law2016,Yan2016survey,yan16calibration,Law2021}. After the MaNGA data are reduced, they are run through a Data Analysis Pipeline \citep[DAP,][]{Westfall2019,Belfiore2019}, which outputs maps of various quantities. In this work we make extensive use of the {\it Marvin} visualisation and data access tools designed for use with MaNGA data and maps \citep{Cherinka2019}. 
 
\begin{table*}
\caption{\label{tab:samples}Galaxy samples and selection in GZ:3D. We release masks from the final phases only, as we only recommend their use for science. The number are MaNGA targets shown to GZ:3D volunteers for each workflow task, not the  number which have successful masks or clusters. The total DR17 sample size will be 10,010 unique galaxies with high quality data cubes (SDSS Collaboration et al. in prep.); the number in this table is the number also in GZ:3D.}
\begin{tabular}{lcccc}
Task & DR14  & All MaNGA targets  & DR17$^*$  & GZ2 Pre-selection \\
& (Phase 1) & (Phase 2) & (Subset of Phase 2) & \\
\hline
Galaxy Centre & 2778 & 29831 & 9188 & All \\
Foreground Stars & 2778 & 29831 & 9188 & All \\
Bars & 175 & 5456 & 1355 & $N_{\rm bar} > 0.2N_{\rm tot}$\\
Spirals & 294 & 7418 & 1973 & $N_{\rm 1-4} > 0.2N_{\rm tot}$ \\
\end{tabular}
\end{table*}

The targeting files presented in \citet{Wake2017} list roughly three times as many potential MaNGA galaxies as were planned to be observed to complete the sample of $N\sim10,000$. At the time this project was designed, it was unclear exactly which MaNGA target galaxies would ultimately be observed, so MaNGA target galaxies were considered as the input sample for {\it Galaxy Zoo: 3D}, and masks for 29,813 galaxies are released for use. MaNGA finished observing in 2020, with a final total of 10,010 high quality data cubes with unique MaNGA-IDs which will be released in SDSS-IV Data Release 17 (DR17, SDSS Collaboration in prep.). The majority of these (9,188, or 92\%) have {\it Galaxy Zoo: 3D} analysis. Detailed Galaxy Zoo (GZ2) classifications \citep{Willett2013} are available for most of the MaNGA target galaxies (these data were released as a Value Added Catalog, or VAC in DR15, \citealt{DR15}\footnote{Available at \\ \url{https://www.sdss.org/dr16/data_access/value-added-catalogs/?vac_id=manga-morphologies-from-galaxy-zoo}}), and this was the input sample used to create the GZ:3D list. MaNGA galaxies were classified in {\it Galaxy Zoo} (mostly GZ2) across a range of phases, and some are missing due to position errors, or other catalog issues. We decided to only run GZ:3D on galaxies which had GZ2-type morphologies available, resulting in 822 observed  MaNGA galaxies with no GZ:3D masks (128 of these are MaNGA filler, or ancillary targets not in the original targeting file, and 434 have since been classified in Galaxy Zoo\footnote{An update to the MaNGA-Galaxy Zoo match will be released as a VAC in DR17}, but not yet GZ:3D).  The benefit of limiting the sample to galaxies with GZ2 classifications, was that in planning GZ:3D we were able to ask volunteers to identify the region covered by bars and spirals only in those galaxies in which GZ2 volunteers had marked such features visible (the numbers of which are shown in Table \ref{tab:samples}).
Furthermore, volunteer testing indicated that drawing spiral arms on very flocculent spirals (5+arms in the GZ2 classifications) was prohibitively hard, so only galaxies with spirals with $n \leq 4$ arms were included in GZ:3D. In this project, we cared about what volunteers could {\it see} to be able to draw, rather than the real underlying morphology of a galaxy, so we used raw vote counts (not debiased to account for the blurring effects of distances, as described by \citealt{Willett2013,Hart2016}), and because we did not want to risk missing galaxies with bars or spirals, we used conservative cuts of $N_{\rm bar} > 0.2N_{\rm tot}$, where $N_{\rm bar}$ is the number of volunteers who reported seeing a bar, and $N_{\rm tot}$ is the total number of volunteers who saw the galaxy in GZ2 (typically around 50) and $N_{\rm 1-4} > 0.2N_{\rm tot}$, where $N_{\rm 1-4}$ is the sum of the number of volunteers who reported seeing 1, 2, 3 or 4 spiral arms. These cuts are aimed to make a complete, but not a clean sample of galaxies with visible bars (this cut should pick up both weak and strong bars; \citealt{Geron2021}) and spirals, but will miss the most flocculent spiral types. 

For the first phase of the project, we used a sub-sample of MaNGA target galaxies -- those which had been observed and released in the SDSS-IV DR14 \citep{DR14} release of MaNGA (internally known as MaNGA Product Launch 5 - MPL-5). For the the second (and final) phase, we used the entire target sample \citep{Wake2017}. A summary of the sample sizes, including the numbers of galaxies with GZ:3D masks in the final DR17 MaNGA sample, is provided in Table \ref{tab:samples}. We recommend the use of only the final GZ:3D masks (from the second and final phase) but discuss data from the first phase because it was used to inform decisions on the final phase. 

\begin{figure*}
\includegraphics[angle=0,width=15cm]{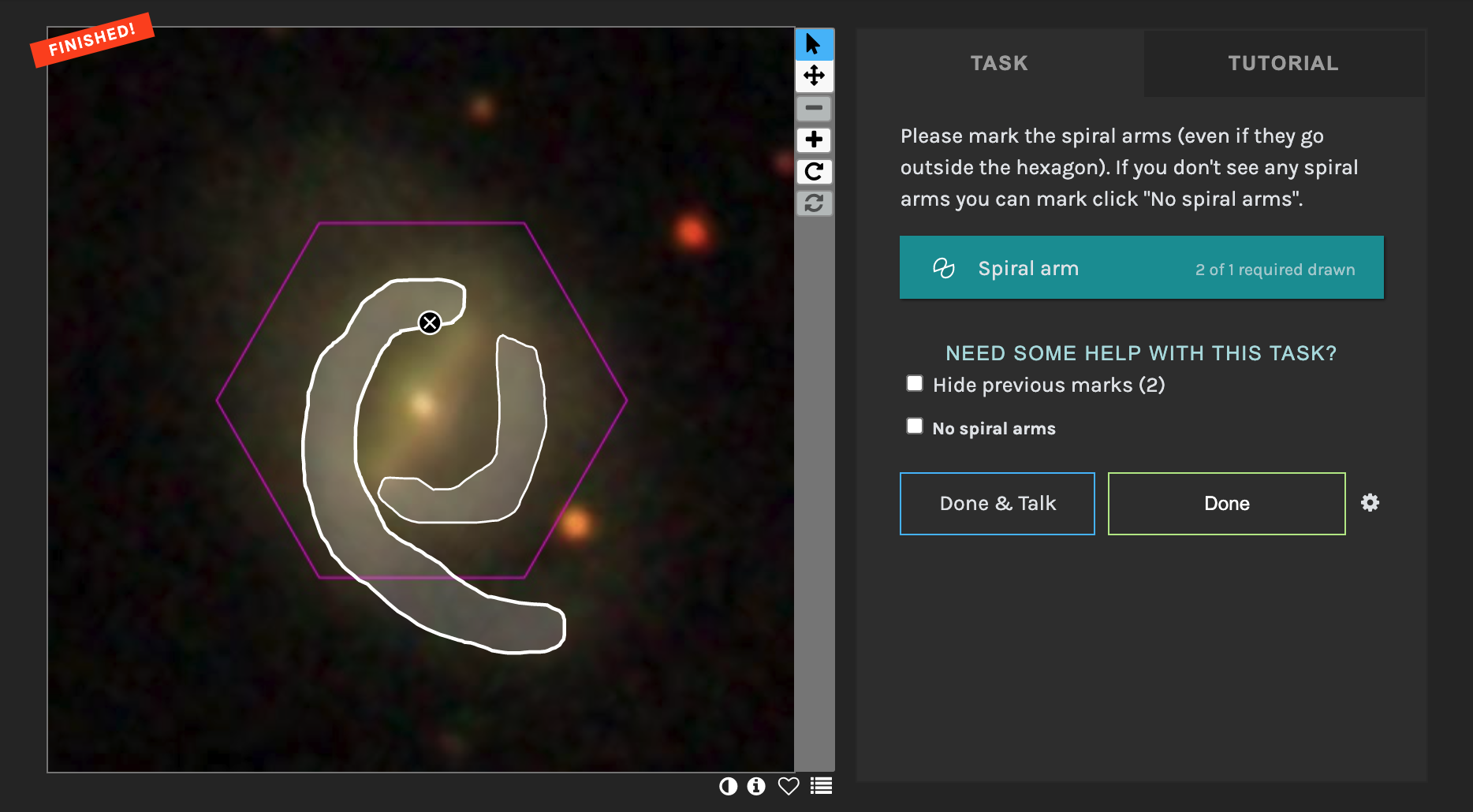}
\caption{A screenshot of the spiral arm classification page from the second phase of the project (after all drawings had been collected), with an example spiral drawn. This is MaNGA-ID 1-63390.}
\label{fig:spiralarmpage}
\end{figure*} 

\subsection{Generating Images}
Custom cutouts of $gri$ images from SDSS-I/II legacy imaging \citep{Gunn1998,York2000,Strauss2002} with an overlay of the approximate location of the MaNGA hexagon (representing the hexagonal fiber bundle; \citealt{Drory2015}) were generated. SDSS images have median resolution of 1.3\arcsec ~(in $r$-band\footnote{\url{https://www.sdss.org/dr16/imaging/other_info/\#SeeingandSkyBrightness}}). GZ:3D images were selected to have a pixel scale of 0.099 arcsec/pixel and were generated to have a field of view of 52\arcsec ~(525$\times$525 pixels), selected to be twice the size of the largest MaNGA bundle on the sky. 

\subsection{Building the Site}\label{sec:building}

The GZ:3D project website\footnote{\url{https://www.zooniverse.org/projects/klmasters/galaxy-zoo-3d}} was built using The {\it Zooniverse Project Builder} software,\footnote{\url{https://www.zooniverse.org/lab}} which allows users to build simple {\it Zooniverse} citizen science projects via a browser interface. Four tasks (or workflows) were developed: (1) clicking on the image to identify the location of one or more galaxy centres; (2) clicking on the image to identify the location of one or more foreground stars; (3) drawing a box around the region of a galactic bar; and (4) a free-hand spiral arm drawing tool to draw around spiral arms. The classification interface for spiral arm marking from the second phase is shown in Figure~\ref{fig:spiralarmpage}. The interface from the first phase, and for bar drawing and for galaxy centre/foreground star marking were similar. 

Volunteers had access to a tutorial to guide them through each task, and were able to select which task (or workflow) to participate in. A decision was made to reduce cognitive load by asking volunteers to do only one task on each galaxy (galaxy centres/foreground stars together) at a time. In the first phase, volunteers were asked to identify features only inside the hexagon on the $gri$ colour image of the galaxy; this instruction was modified in the second phase (as is discussed below). 

\begin{figure*}
\includegraphics[angle=0,width=15cm]{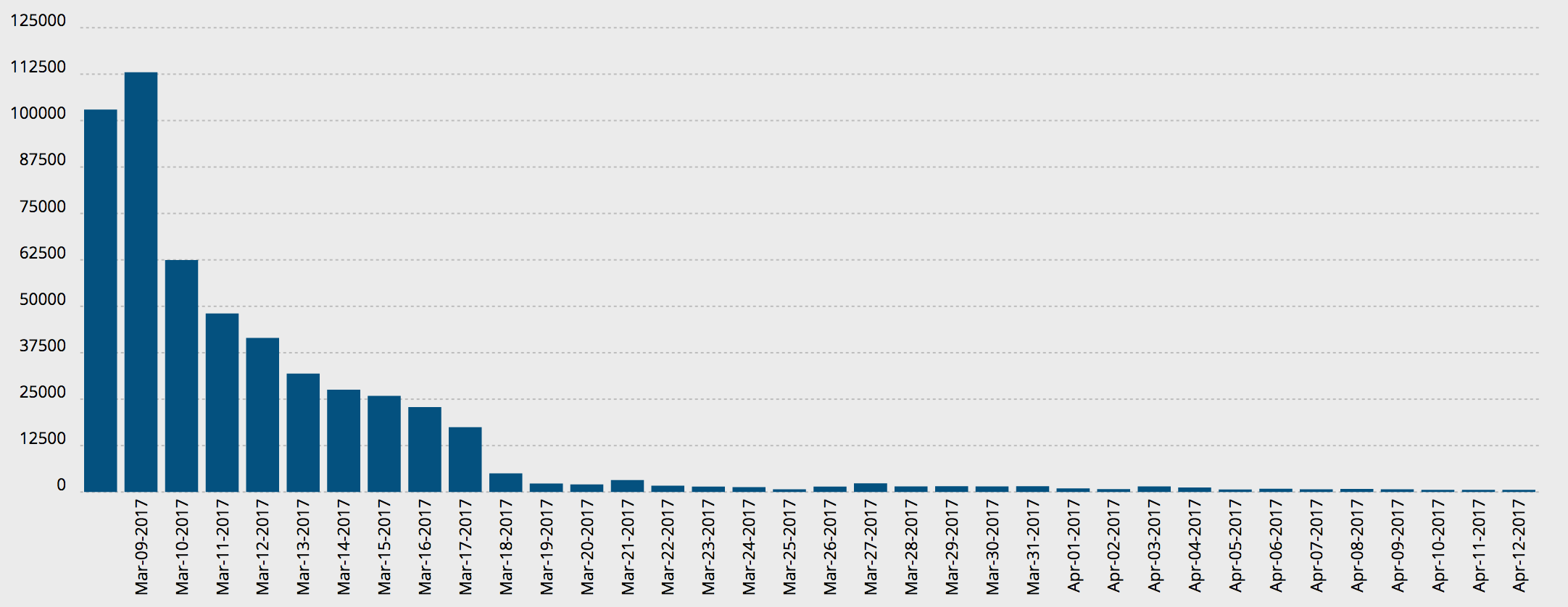}
\includegraphics[angle=0,width=15cm]{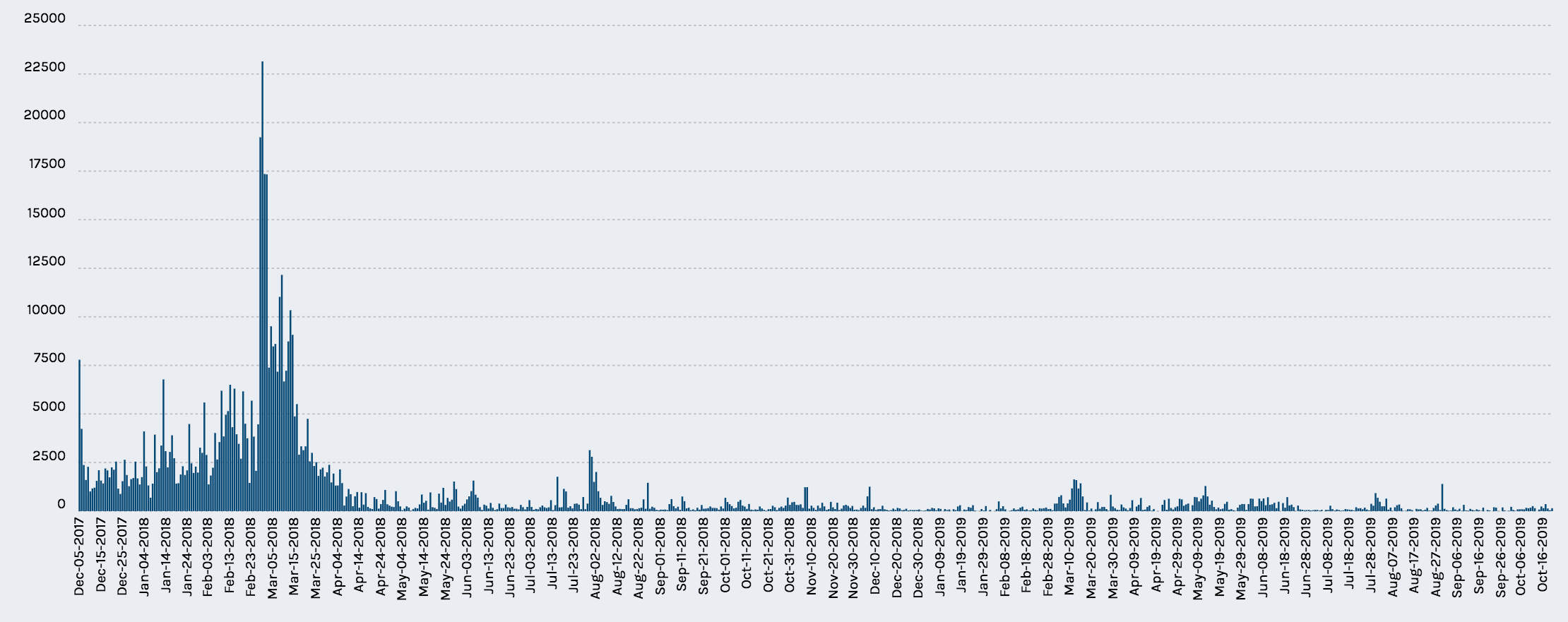}
\caption{Upper: The total number of classifications recorded by the project by day from launch until 12th April 2017. The galaxy centre/foreground star marking task was completely on 17th March 2017. Lower: the total number of classifications in the second phase of data collection from 4th Dec 2017 to 20th October 2019.}
\label{fig:classifications}
\end{figure*}

After a beta test period in the Summer of 2016, the project launched
on 8th March 2017 with a Newsletter sent to the entire Zooniverse volunteer base. 
Figure \ref{fig:classifications}, shows a count of volunteer engagement rates with the site during both this first, and the later second phase. The galaxy centre/foreground star marking workflows attracted significantly higher rates of engagement than the bar or spiral arm drawing tasks. Galaxies were retired from a specific workflow when 15 volunteers had made marks for that task. All first phase galaxies were retired from the galaxy centre/star marking task on 17th March 2017, resulting in a notable drop of volunteers on the project after that day (see Figure \ref{fig:classifications}), when only the more challenging bar and spiral arm drawing tasks remained. 

After some analysis had been done with the initial subject set of DR15 MaNGA galaxies, it was noticed that the hexagons indicated on the initial subject set were slightly too small. This was caused by an error in interpreting hexagonal diameters -- the first phase images included a hexagon that was a factor of $2\sqrt{3}/5$ too small - the difference between corner-to-corner versus edge-to-edge sizes. In addition, some of the MaNGA target galaxies did not have GZ2 classifications available in \citet{Willett2013} at the time of launch and had been added as a small set to the main {\it Galaxy Zoo} site.\footnote{\tt www.galaxyzoo.org} Other small changes were made to improve the classification experience: 
\begin{itemize}
    \item Galaxies with less obvious (i.e. lower cuts on GZ2 morphology votes) bars and spiral arms were included for feature marking. 
    \item A button was added to indicate if an image does not contain a galaxy/bar/spiral arms. 
    \item We asked volunteers to identify features both inside and outside of the hexagon, partly to catch foreground stars just outside which can impact the light which is collected by MaNGA, but also because the drawn hexagon (even at the correct size) is an approximation of where MaNGA bundles collect light in the dither pattern described in \citet{Law2015}. 
\end{itemize}

For the second phase of GZ:3D we included all MaNGA target galaxies, including repeating those previously classified in the first phase. This second phase launched to previously registered GZ:3D volunteers on 4th December 2017, and classifications were collected at a more modest rate than in the first phase (see the lower panel of Figure \ref{fig:classifications}). A boost up to a peak rate of roughly 23,000 classifications per day was obtained by emailing {\it Galaxy Zoo} volunteers on 27th Feb 2018. Following this, foreground stars and galaxy centres were completed on 5th April 2018, and bar masks were completed in early August 2018. A post to the {\it Galaxy Zoo} blog\footnote{\tt{https://blog.galaxyzoo.org}} brought another small spike in classifications, after which spiral masks took until October 2019 to complete at a modest rate of 500-1000 classifications per day. 

\subsection{Generating Masks}

In GZ:3D, every task was retired when it had been done by 15 volunteers, although sometimes the total number recorded is larger than 15. All classifications are combined to created a single consensus classification for each task. 

For the galaxy centre, and foreground star marking tasks, a point clustering algorithm (Density-based spatial clustering of applications with noise or {\tt DBSCAN}, \citealt{DBSCAN}) was used.\footnote{Specifically the {\tt scikit-learn} implementation available at\\ \url{https://scikit-learn.org/stable/about.html\#citing-scikit-learn}} A cluster of points (or volunteer clicks) was defined as a group of three or more points within five pixels (or 0.5\arcsec) of each other. The mean and covariance of these points within a cluster was then used to create both a centre and confidence region (a 2$\sigma$ region mask). The point count is provided and can be used a a measure of confidence in the cluster. 

For the bar and spiral arm drawing tasks, ``count masks" were created for each pixel in the image. These record the number of volunteers who included that pixel inside of their bar or spiral arm drawing respectively. One complication with this technique, for the free-hand spiral arm drawing task, is that volunteers could draw shapes which were self-intersecting polygons (e.g. a five-pointed star drawn with a single line). Some of these shapes do not have a well defined ``inside" so were removed from the aggregation, this, as well as volunteers not completing a task can result in a small number of galaxies with fewer than $N=15$ classifications for the spiral mask. 

\begin{figure}
\includegraphics[angle=0,width=8cm]{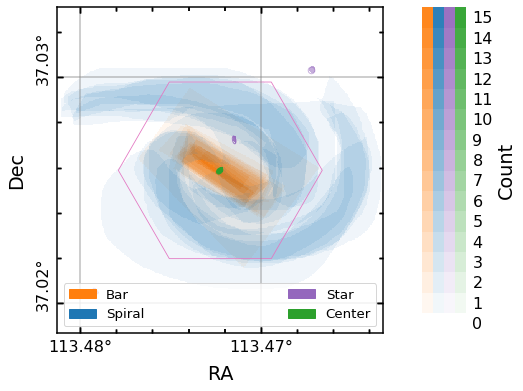}
\caption{An example of the masks drawn on a galaxy in GZ:3D. This shows all kinds of masks with colour indicating the type (green for galaxy centre, purple for star, orange for bar and blue for spirals), and the density indicating the number of volunteers who created the mask in that pixel. Their are two foreground stars identified in this example, one is outside of the MaNGA bundle. This galaxy is MaNGA-ID 1-604761. }
\label{fig:examplemask}
\end{figure}

An example aggregation mask which includes a bar, spiral, foreground star (in this example outside the bundle) and galaxy centre is illustrated in Figure \ref{fig:examplemask}. The code used to perform the aggregation, and to make this plot for any galaxy in the GZ:3D sample has been made available\footnote{\url{https://github.com/CKrawczyk/GZ3D\_production}}. The colour bar indicates which colour corresponds to which feature, with the density of the colour indicating the volunteer count (i.e. the most opaque colours are reserved for spaxels or points marked by all volunteers). This gives a sense of how a scientist using GZ:3D masks must decide on a count threshold (in number, $N$, or fraction, $f$ of answers) which will define a specific region (see Section \ref{sec:thresholds} below). 

\section{Results}\label{sec:results}

\subsection{Galaxy centres}\label{sec:centres}
 
\begin{figure*}
\includegraphics[angle=0,width=15cm]{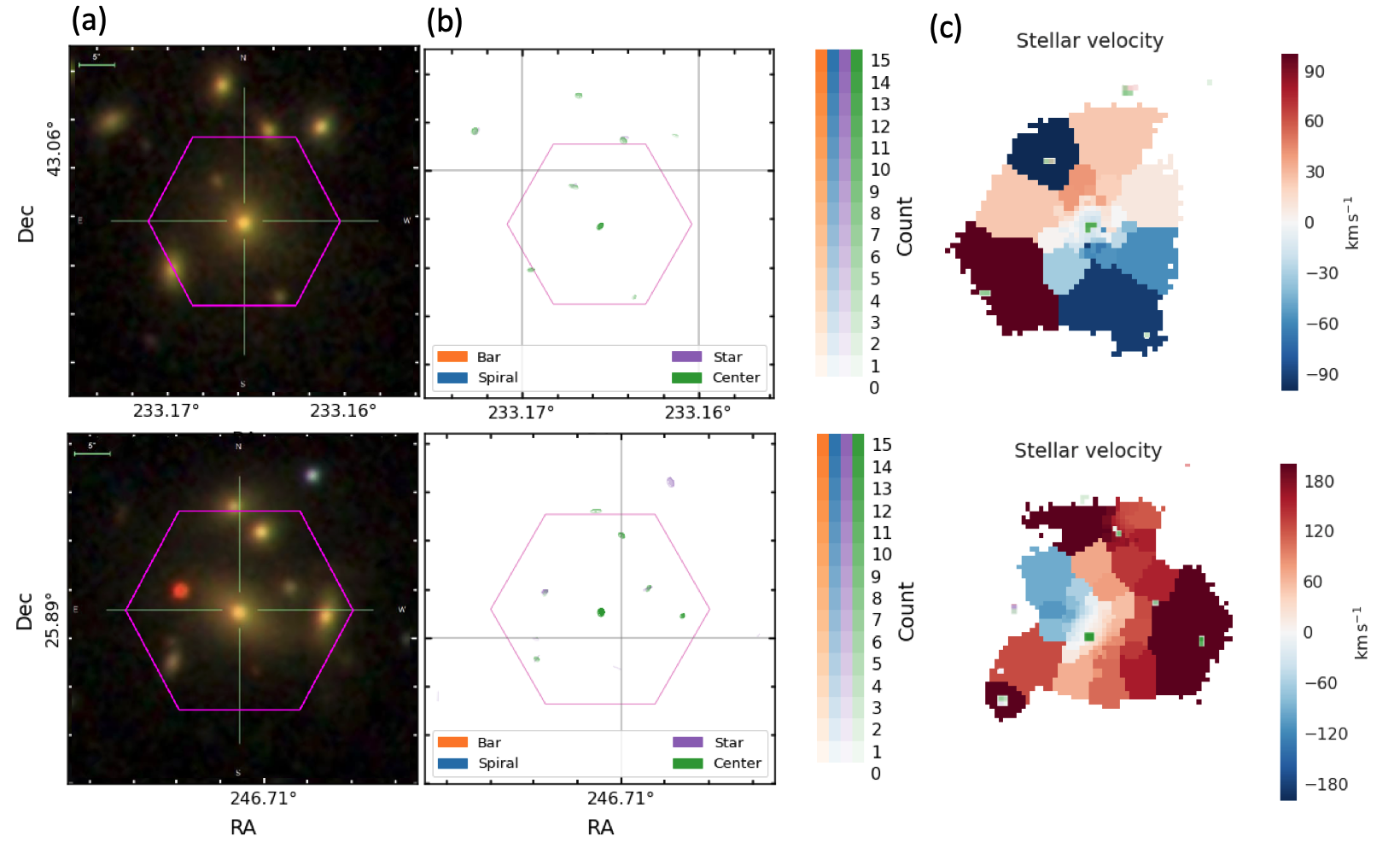}
\caption{Examples of MaNGA targets in which GZ:3D volunteers marked large numbers of galaxy centres (in these examples five and six GZ:3D identified galaxy centre clusters within/close to the bundle). The upper row shows MaNGA-ID 1-199224, while the lower row shows MaNGA-ID 1-295355. Column (a) is the optical image with the hexagon overlaid; (b) are the GZ:3D masks, showing multiple galaxy centres (and in MaNGA-ID 1-295355 two stars; one inside the bundle). Column (c) shows the stellar velocity with spaxels tagged as galaxy centres overlaid. }
\label{fig:centres}
\end{figure*}

The marking of galaxy centres by GZ:3D volunteers can be used to identify MaNGA galaxies with a significant offset from the centre of their bundle and also MaNGA bundles which include multiple galaxies (for an example see Figure~\ref{fig:centres}). One interesting application could be finding offset bars (bars offset from the photometric centre; \citealt{Kruk2017}). 
  
In total we provide information for 36,293 clustered galaxy centres (i.e. clustered points in or near the bundles) in the field of view of objects in the MaNGA target file. These were all marked by $N>3$ volunteers; 85\% of these clusters have $N>6$ or typical $f>40\%$. Of these, 11,161 centres are in targets observed by MaNGA (9841 with the higher threshold). 
  
We find 3200 MaNGA galaxies with multiple galaxy centres (1,819 after applying the $f=40\%$ threshold). Two examples of MaNGA targets with eight clustered galaxy centres are shown in Figure \ref{fig:centres}. 
  
In bundles with only a single galaxy centre identified, we find good agreement between the position and the catalogued galaxy position, with offsets having a mean value of 0.34$\pm$0.16\arcsec, which is well within a single MaNGA spatial resolution element (2.5\arcsec ~after the cube is reconstructed, \citealt{Law2016}). Some bundles were deliberately offset during initial test observations, or as part of an ancillary program to cover merging galaxies in MaNGA. The largest offset we find in this sample with single centres IDs (with an $f=40\%$ threshold) is 3.4\arcsec, which reveals an object (MaNGA-ID 1-38770) with a significant tidal tail. Some of the deliberate offsets were larger than this.

\subsection{Foreground Stars}\label{sec:stars}

Foreground stars are present in a small fraction of MaNGA bundles and (if unflagged) cause problems for the MaNGA DAP. To solve this problem, foreground star masks are included in the Data Reduction Pipeline \citep[DRP,][]{Law2015} output which forms the input to the DAP. The foreground star masks used in both DR13 \citep{DR13} and DR14 \citep{DR14} MaNGA data were created by MaNGA team members visually inspecting all galaxies observed in the first 2 years of MaNGA operations.\footnote{Foreground star visual inspection was mostly done by Karen Masters and Lin Lin.} However in DR15 \citep{DR15} and later releases, these star masks were supplemented by the masks created via GZ:3D. 

The MaNGA team foreground star identification listed 277 stars identified in 260 individual MaNGA galaxies in the DR14 (or MPL-5; \citealt{DR14}) subset. This provided a comparison set for GZ:3D foreground star identifications from the first phase. Considering foreground stars identified by five or more volunteers only, GZ:3D identified 47\% of the foreground stars in the team list and 23\% of the GZ:3D identified stars were found in the MaNGA team list (within 5 arcsecs). We expect that GZ:3D would find stars missed by the team, since individual human classification suffers from a level of human error which crowdsourcing solves. The majority of the foreground stars in the team list which were not in GZ:3D (at the $N>5$ count threshold) were outside or around the outskirts of the hexagonal bundle. We suggest these missing stars were primarily caused by the instructions in the first phase: to mark stars only within the hexagon, combined with a displayed hexagon which was slightly too small as explained previously. This formed part of the motivation to classify all MaNGA target galaxies again in Phase 2 (see Section \ref{sec:building}), and to modify the instructions to request stars both inside and outside the bundle to be marked. This adjustment improved significantly the fraction of team identified stars found in GZ:3D, with 80\% identified (at $N>5$). 

\begin{figure*}
\includegraphics[angle=0,width=16cm]{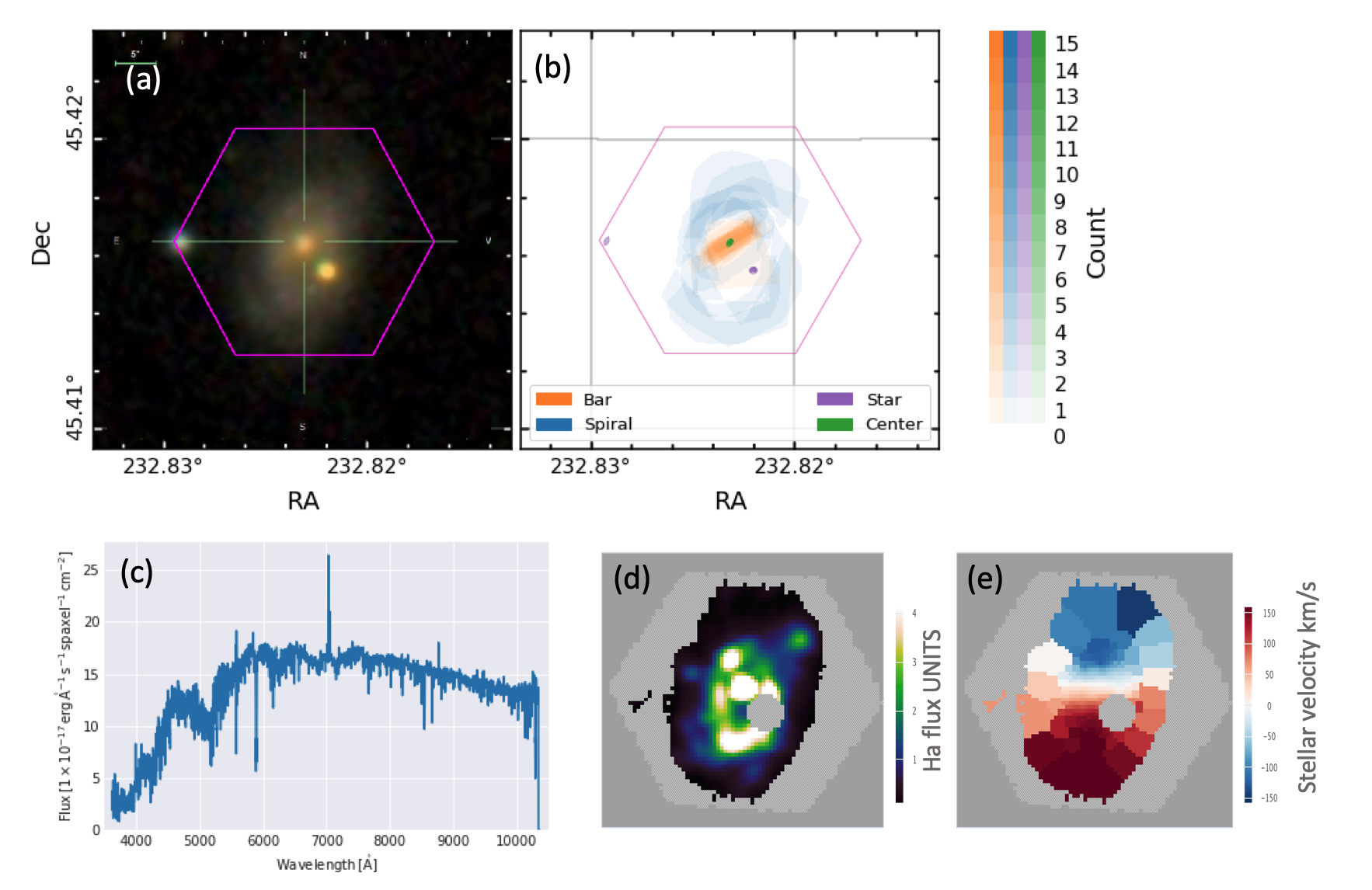}
\caption{An example of a MaNGA galaxy (MANGA-ID 1-247456)  with two foreground stars identified by GZ:3D. Shown are (a) the optical $gri$ image; (b) the GZ:3D classification mask;  (c) the spectrum of the most central star which was identified by 13/15 GZ:3D volunteers and shows a spectrum characteristic of a K-type dwarf star (having a strong Mg line at 5200\AA ~and Na line at 5800\AA) which, based on {\it Gaia} parallax is just 1.1kpc from the Sun; (d) a H$\alpha$ emission map from {\it Marvin}; and (e) a stellar velocity map from {\it Marvin} both with the central foreground stars masked.}
\label{fig:star}
\end{figure*}

 For MaNGA galaxies which have been observed, it is possible to extract spectra at the position of a marked  ``foreground star" in order to check if it is actually a star, or is some other kind of points source (e.g. a compact HII region or background quasar). We did this for all marked stars (regardless of the number of volunteers who marked them) in the MaNGA DR15 sample (internally known as MPL-7) which was available internally in June/July 2018 \footnote{KNAC funded REU project of Daniel Finnegan}. We made use of {\it Marvin} \citep{Cherinka2019} to extract averaged spectra from a circle with radius 2.5\arcsec (the typical MaNGA PSF) at the clustered location of point sources marked as foreground stars. This resulted in a return of 2037 spectra, an example is shown in Figure \ref{fig:star}. This exercise revealed that all point sources identified by $N\geq10$ volunteers are stars (a private {\it Zooniverse} project was used to ease the process of classifying the spectra from marked locations), while at lower volunteer thresholds some HII regions and other compact bright regions (sometimes background or foreground galaxies) were identified (with $N\geq6$ we found 80\% of objects were stars, and for $N\geq3$, 50-60\% were stars). In order to be conservative and not exclude spaxels of interest from the DAP analysis, we therefore make use of a threshold of $N\geq10$ to identify foreground stars which are flagged in the DAP \citep{Westfall2019}. In the entire MaNGA target list this identified 3658 foreground stars, though many of these will be in target galaxies never observed by MaNGA. Of the 11273 data cubes in MPL-11 (the final internal release which will become the DR17 sample, of which 10,010 are unique MaNGA-IDs with high quality data), 1085 (9.6\%)  had foreground star flags marked. 

\subsection{Bar Regions}
\begin{figure*}
\includegraphics[angle=0,width=15cm]{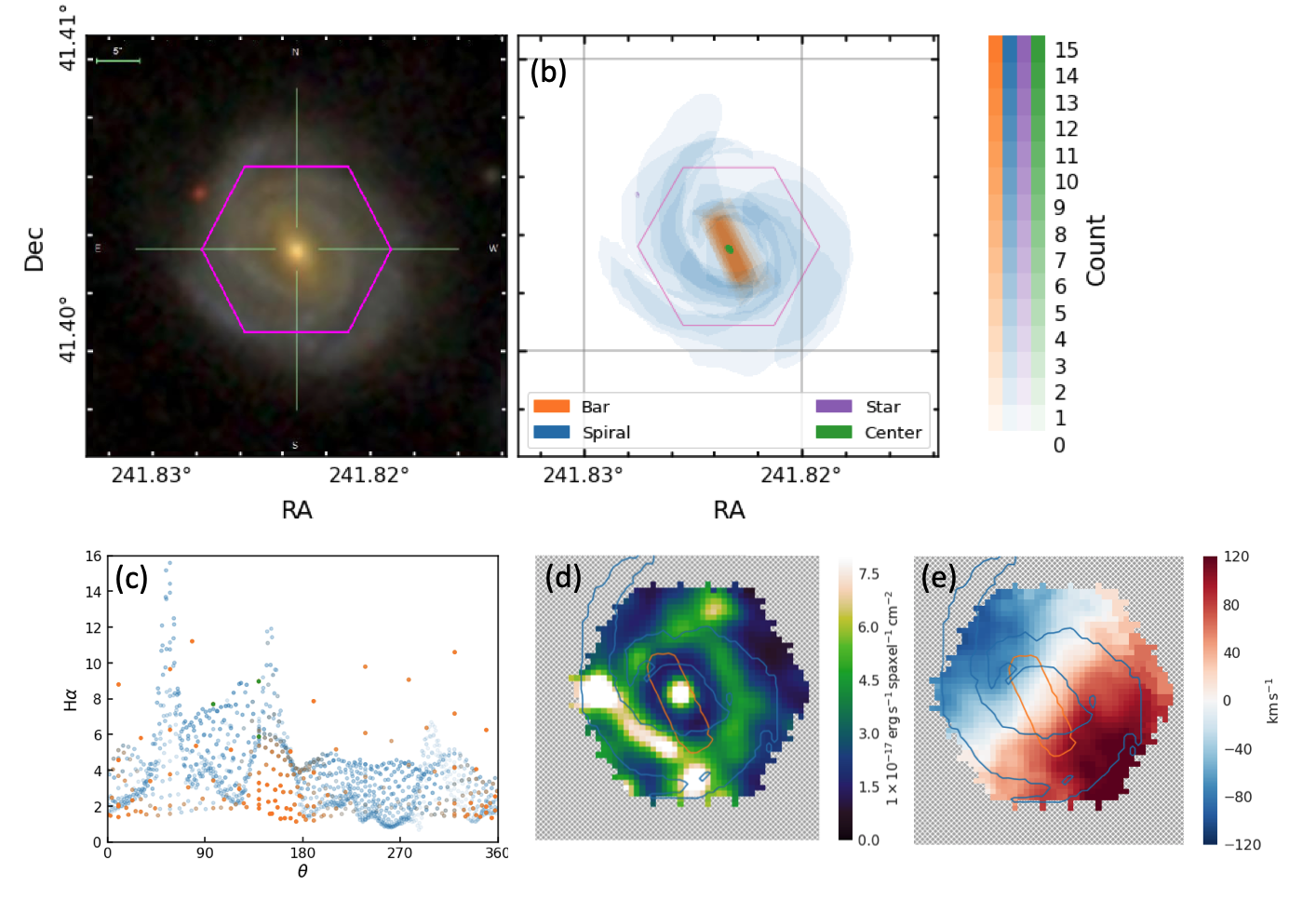}
\caption{An example of a MaNGA galaxy (MaNGA-ID -248420) with a GZ:3D bar mask and spiral mask. Panel (a) shows the image shown to {\it Galaxy Zoo: 3D} volunteers while the pixel masks output are shown in panel (b). Panel (c) shows a scatter plot of the H$\alpha$ flux in each spaxel, as a function of azimuthal angle. Points are colour coded by the GZ:3D identification (e.g. orange for spaxels in the bar, blue for spiral arm spaxels). This shows how the peak in H$\alpha$ emission varies significantly with angle. The bar and spiral masks are used to create the overlays on the MaNGA H$\alpha$ emission map shown in panel (d), and stellar velocity map in panel (e). The bar contour is at the 7/15, or 47\% threshold level, while the spiral is at the 3.5/15, or 23\% threshold. }
\label{fig:barspiral}
\end{figure*}

 Galactic scale bars are present in a large fraction of disc galaxies (from 30-60\% depending on the wavelength of observation, and the technique used to identify these linear features, e.g. \citealt{MarinovaJogee2007,Sheth2008,Masters2011bars}). Since the strong correlation between optical colour (as an indicator of quenching) and bar fractions was first noted \citep{Masters2011bars} there has been significant interest in the role bars may play in the quenching of star formation in disc galaxies \citep[e.g.][]{Masters2012,Kruk2018,Gavazzi2015,Khoperskov2018,George2019,Maeda2020}. Even before this, the role of bars in radial mixing (and corresponding impact on radial gradients) had been much debated (see \citealt{Zurita2021a,Zurita2021b} for a discussion of the history of this topic). Bars, particularly strong bars which extend over a significant fraction of the galaxy disc, have a strong directionality (that's what makes them a bar!), so azimuthal averaging in galaxies with strong bars is highly likely to oversimplify and obscure radial gradients \citep{SanchezBlazquez2014}. The interpretation of radial gradients in galaxies is already complex, revealing as it does the intricate interplay between star formation rates, intergalactic gas inflow and radial migration driven by various internal processes. Hence, the identification of bar regions to help disentangle the impact of bars on azimuthally averaged gradients from MaNGA data was one of the primary motivations for the GZ:3D project. An example bar mask from the project is shown in Figure \ref{fig:barspiral} which shows the optical image, mask from the volunteers, and details of the H$\alpha$ flux and stellar velocity from MaNGA with GZ:3D identified bar overlaid. 

 The bar masks from GZ23D have already been used in a number of studies with MaNGA data. The first published result was \citet{FraserMcKelvie2019}, who used bar masks to investigate gradients in stellar populations both along and outside bars at the same galactic radii, finding that there are indeed clear differences - there were flatter age gradients in bars than in the inter-bar disk regions at the same radius. \citet{FraserMcKelvie2020} went on to use MaNGA to reveal how H$\alpha$ morphology correlates with bars. They found that in the small fraction (18\%) of bars in their sample with strong H$\alpha$ in the bar, the ionization source (using the \citealt{BPT}, or BPT method to identify the likely source of ionization) was star formation. This work also demonstrated how the angle of the H$\alpha$ bar leads that in the white-light image (dominated by older stellar populations) by up to 20$^\circ$ in most galaxies; a visual confirmation that star formation is occurring primarily at the leading edge of the bar, in line with theoretical predictions \citep[e.g.][]{Renaud2015}. Measurements of this offset would provide a way to measure the pattern speed of bars [e.g. as was done by \citet{Peterken2019Density} for a spiral arm]. 
 
\citet{Peterken2019TimeSlice} used both bar and spiral masks from GZ:3D for the galaxy MCG+07-28-064 to demonstrate a new technique of reconstructing spatially resolved star formation histories from MaNGA data cubes. The bar (and spiral arm) masks were used as a visual indication of where these features are relative to features in stellar population metallicity as a function of age. 
 
 \citet{Krishnarao2020} have used GZ:3D bar masks in a sample of Milky Way analogues with bars to identify the source of ionization in the bar itself and in the inter-bar region at the same radius. They discovered LI(N)ER like emission is more common in the inner region (inside the bar radius) in barred galaxies than it is in the outer region (a finding also reported by \citet{Percival2020} in a much smaller sample), and that there is often a ring of star formation at the end of the bar. In Appendix A of \citet{Krishnarao2020}, there is also a published test of bar lengths measured using GZ:3D masks compared with those measured using the Fourier decomposition method described in \citet{Kraljic2012}. This result demonstrates that as the GZ:3D vote threshold is increased, GZ:3D bars get shorter such that the agreement between the two measures of bar length improves, but at the cost of a smaller fraction of galaxies with bars that have usable GZ:3D bar masks. In a nice demonstration of the kind of considerations users of GZ:3D output need to make, \citet{Krishnarao2020} concluded that a vote threshold of 20\% was optimal for their application (See Section \ref{sec:thresholds} below for further discussion of GZ:3D vote threshold choices). 
 
 D. Krishnarao et al. in prep. have gone on to use GZ:3D bar masks to identify dark gaps in stellar surface density (and $g$-band imaging) which they link to orbital resonances generated by the bar. This measurement, which they calibrate with similar measurements in an N-body simulation, allows them to identify the co-rotation radius of the bars, and estimate pattern speeds. This work does additional testing of GZ:3D bar lengths, comparing them to \citet{Guo2019} and finding good agreement at the 40\% threshold level. 

\subsection{Spiral Arms}

 Spiral arms are a common feature in disc galaxies. They range from the highly dominant, usually two-armed ``grand design" spirals to the more flocculent (irregular and patchy) types, which tend to have large numbers of arms which are hard to separate (see \citealt{Hart2016} for a summary of the demographics of spiral arms as observed in Galaxy Zoo.)
 
  For the GZ:3D project, we invited volunteers to draw spiral arms in MaNGA target galaxies where Galaxy Zoo 2 \citep{Willett2013} volunteers had indicated they could see four or fewer spiral arms. During beta testing, volunteers indicated that the spiral drawing task was hard to complete, particularly where there were large numbers of very flocculent arms. Even with these many armed spirals removed, it was spiral arm drawing which took the longest to complete (see Section 2.3). 
  
   An example spiral arm drawing and mask, and plots showing the H$\alpha$ and stellar velocity maps from MaNGA with the outline of a spiral mask (with the threshold at 3.5/15 volunteers or 23\%) is shown in Figure \ref{fig:barspiral}.

 The MaNGA team have already made some use of spiral arm masks in published results. As noted above, \citet{Peterken2019TimeSlice} used the spiral mask to illustrate features in MCG+07-28-064; in addition, \citet{Peterken2019Density} used the spiral arm mask to indicate where the spiral arms are in UGC 3825. In this galaxy, the MaNGA maps of H$\alpha$ and young stars were cross correlated, revealing the stellar population age gradient signature expected for a density wave spiral (with H$\alpha$ tracing star formation happening in the spiral, and young stars ``behind it"). 

In a paper about the dust content of galaxies as revealed by gas and stellar tracers, \citet{Greener2020} used spiral masks to investigate arm-interarm differences in dust content. No differences were seen, which is surprising, as generally dust lanes are associated with spirals. This led to speculation that the 2.5$\arcsec$ spatial resolution of MaNGA data was blurring out the signal, or potentially that dust lanes are excluded from drawn arms in GZ:3D.  

\subsection{Star Formation in Spirals}\label{sec:spiralSF}

In this section we present a new result making use of spiral arm masks from GZ:3D to investigate the enhancement of star formation linked to spirals.

\begin{figure}
\includegraphics[angle=0,width=8cm]{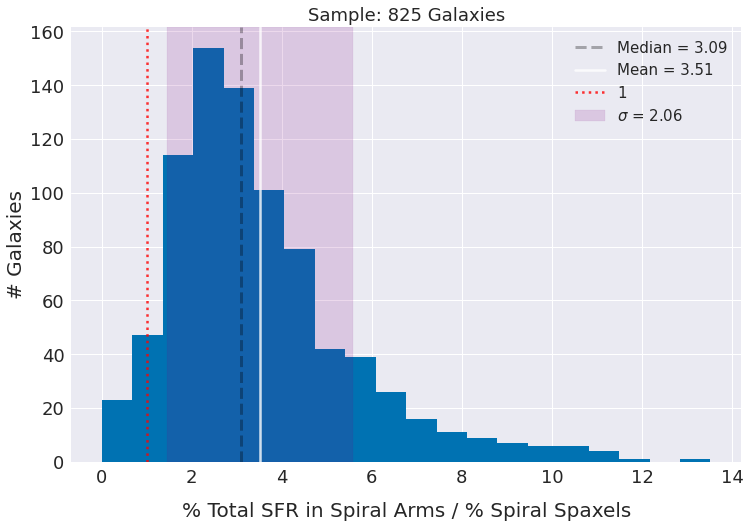}
\caption{The fraction of star formation found within the spiral masks, normalized by the fraction of the area those masks cover. in a sample of 825 MaNGA galaxies with MPL-10 data and GZ:3D spiral masks. If there were no enhancement of SF in the arms we'd expect this to peak at 1.0 (the red dotted line); this measurement shows a peak at a median of 3.1, with $\sigma=2.1$, showing that SF is enhanced in spiral arms. }
\label{fig:SFspiral}
\end{figure}

Spiral arms have long been predicted to trigger star-formation via shocks to the gas clouds \citep{Roberts1969}, this might also be caused by an increased in gas density driven by spiral arms \citep{Kim2020}; in this recent simulation it was estimated that 90\% of star formation happens in spirals arms due to the increased gas density. Here, we make use of the GZ:3D spiral masks to measure the fraction of star formation found in the spiral arms in MaNGA galaxies. We use data from the MaNGA MPL-10 internal release\footnote{MPL-10 was an internal data release within MaNGA which containing data cubes for 9269 unique galaxies - see Table 1 of \citet{Law2021}}. We construct a sample of 825 galaxies with spiral masks (with a spiral threshold of 20\% or 3/15 volunteers marking a spaxel as within a spiral arm) in at least 1.5\% of the spaxels. We  exclude spaxels marks as being within a bar, galaxy centre or star by at least 3/15 volunteers from either the spiral arm, or inter-arm regions.  Star formation estimates are based on the H$\alpha$ luminosity, corrected with the Balmer decrement (using the formula given in \citealt{Spindler2018}). We sum the total SF inside the spiral mask and compare to the area covered by the arms in a range of $0.1 < r/r_e < 1.5$, where $r_e$ is the effective radius of each galaxy. The majority of MaNGA galaxies are covered by the bundle to at least $1.5r_e$. The resulting distribution is presented in Figure \ref{fig:SFspiral}, which shows a peak at a median of 3.1 (standard deviation $\pm2.1$), demonstrating that in this sample, spiral arm spaxels host around 3 times as much star formation as the average spaxel at the same radius.  We alternatively describe this as a star formation enhancement (SFR$_{\rm spiral}$-SFR$_{\rm non-spiral}$)/SFR$_{\rm non-spiral}$, finding an average enhancement of 62\% (median 54\%, and standard deviation 70\%). 

\begin{figure}
\includegraphics[angle=0,width=8cm]{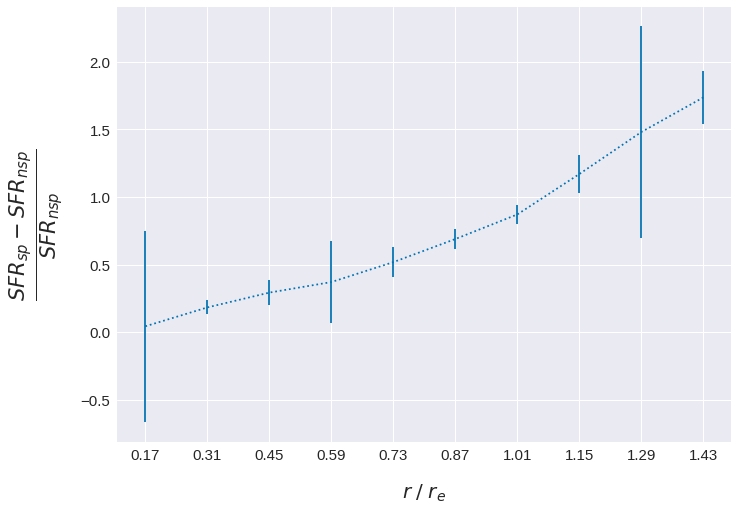}
\caption{The average radial profile (radius scaled to be in unit of galaxy effective radius, $r_e$) of the star formation excess ((SFR$_{\rm spiral}$ - SFR$_{\rm non-spiral}$)/SFR$_{\rm non-spiral}$) found using the GZ:3D spiral masks with a sample of 825 MaNGA galaxies with MPL-10 data. This figure shows how at larger radius, the SFR is particularly enhanced in spiral arms. The errors show the 1$\sigma$ scatter at each radial bin.  }
\label{fig:SFgradient}
\end{figure}

Using the spiral masks, we are also able to construct radial gradients of star formation enhancements linked to spiral arms  (see Figure \ref{fig:SFgradient}). This reveals that at larger radii, star-formation is more strongly enhanced in the spiral arms (from effectively no enhancement at $r<0.2r_e$ to roughly 100\% more SF in the arms at around 1$r_e$). Further analysis of these results in future work will help measure the total contribution of spiral arm enhanced star formation in a large sample of galaxies, and compare the enhancement to gas density enhancements to measure the SF efficiencies.  

\section{How to Use GZ:3D Masks}\label{sec:data}

All GZ:3D results, which cover all MaNGA target galaxies, are published as a Value Added Catalog (VAC) in the Seventeenth Release of the SDSS (DR17; SDSS Collaboration in prep.)\footnote{ \url{https://www.sdss.org/dr17/data_access/value-added-catalogs/}}. 

Each galaxy which has been processed in GZ:3D has one {\tt .fits} file containing the following information:
\begin{itemize}
\item {\tt HDU 0: [image]} SDSS cutout image presented to the volunteer. The pixel scale in these images, and all pixel masks is 0.099 arcsec/pixel (see Section 2.2). 
\item {\tt HDU 1: [image]} Pixel mask of clustering results for galaxy centres. Each identified centre is represented by a 2 sigma ellipse of the clustered points with the value of the pixels inside the ellipse equal to the number of points used belonging to that cluster.
\item {\tt HDU 2: [image]} Pixel mask of clustering results for stars. Each identified star is represented by a 2 sigma ellipse of the clustered points with the value of the pixels inside the ellipse equal to the number of points used belonging to that cluster.
\item {\tt HDU 3: [image]} Pixel mask of spiral arm locations. The values for this mask are the number of polygons describing the spiral drawings overlapping each pixel.
\item {\tt HDU 4: [image]} Pixel mask of bar location. The values for this mask are the number of polygons describing the bar drawing overlapping each pixel.
\item {\tt HDU 5: [table]} Image metadata
\item {\tt HDU 6: [table]} centre cluster data table in both pixel coordinates and RA-DEC. The covariance values used to make the 2-sigma masks are also included.
\item {\tt HDU 7: [table]} Star cluster data table in both pixel coordinates and RA-DEC. The covariance values used to make the 2-sigma masks are also included.
\item {\tt HDU 8: [table]} Raw centre and star classifications provided in pixel coordinates.
\item {\tt HDU 9: [table]} Raw spiral arm classifications provided in pixel coordinates (\ie. the shape of each polygon drawn).
\item {\tt HDU 10: [table]} Raw bar classifications provided in pixel coordinates (\ie. the shape of each polygon drawn).
\end{itemize}

In addition to this, we provide example code to use these files with MaNGA data cubes\footnote{\url{https://github.com/CKrawczyk/GZ:3D\_production}}. This code assumes the user has {\it Marvin} \citep{Cherinka2019} installed and leverages functionality in {\it Marvin} to make working with the MaNGA data cubes or maps and GZ:3D straightforward. It is intended that future releases of {\it Marvin} will include versions of this code.  

A summary file with galaxy meta data (including the GZ2 vote counts used to identify galaxies with likely bars and spiral arms) is also provided, as well as summary files for the galaxy centre identifications, and foreground star identifications. 

\subsection{Mask Levels} \label{sec:thresholds}

 When using crowd-sourced masks (or morphologies) a scientist may choose to use the votes fractions directly as a weighting, but they may also wish to make a choice of threshold in order to create a binary mask (i.e. identifying arms and inter-arm regions, or a region which is associated with the galactic bar). We provide in this section some advice, and examples of threshold choices. We also point the reader to Section 4.3 of \citet{Willett2013} which provides a discussion of similar choices when using Galaxy Zoo morphologies. 
 
\begin{figure*}
\includegraphics[angle=0,width=16cm]{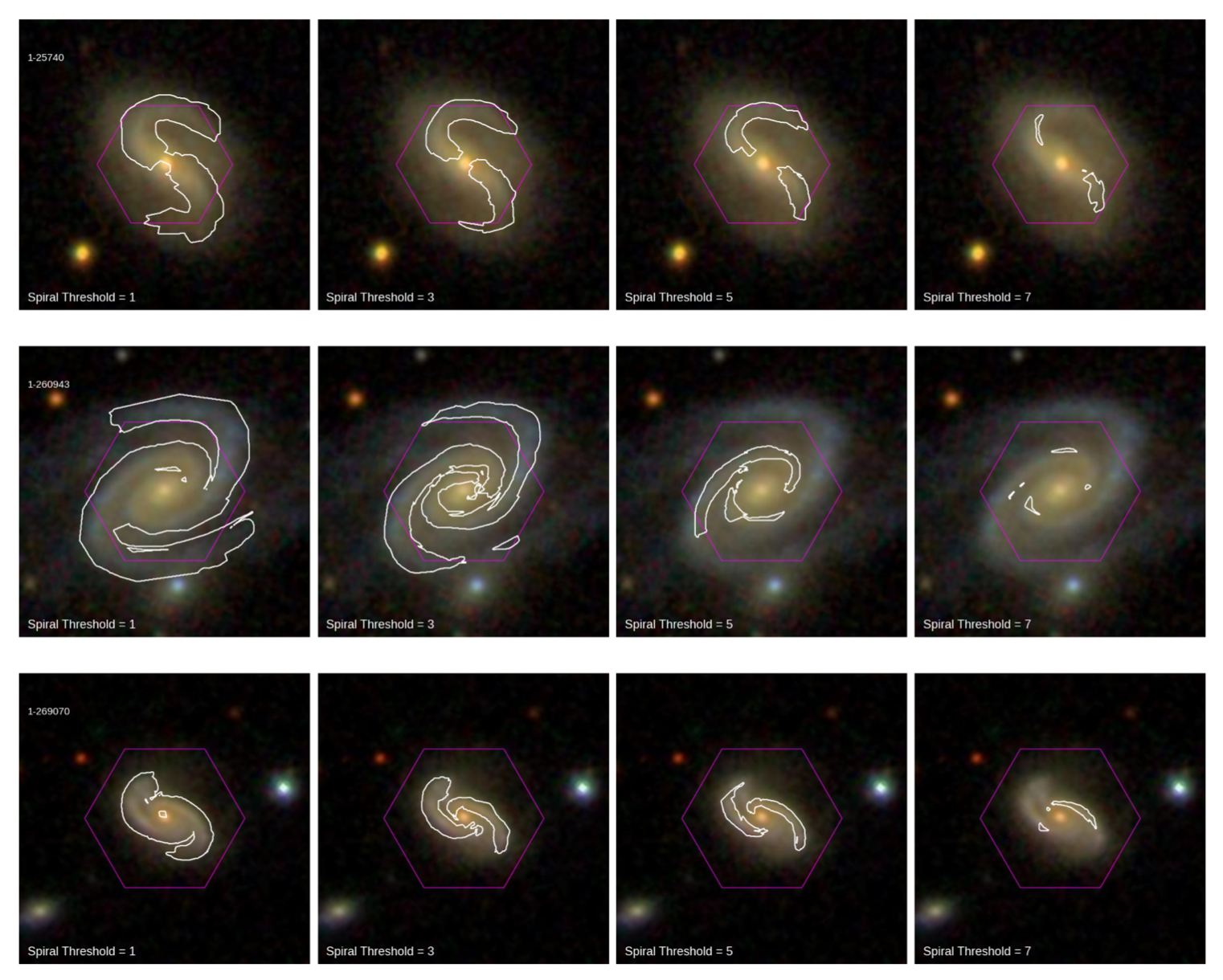}
\caption{Three galaxy images (top to bottom: MaNGA ID: 1-25740, MaNGA ID: 1-260943, and MaNGA ID: 1-269070) displayed with their GZ:3D spiral arm masks overlaid. The masks are shown at different spiral thresholds, from left to right: N = 1, 3, 5, and 7 users mark each enclosed spaxel as part of a spiral. We conclude from this that a threshold of $N=3$ is optimal to identify continuous, but distinct spiral arms.}
\label{fig:spiralthresholds}
\end{figure*}

 The basic trade-off is that the higher the vote threshold, the more confident the user can be in the identification of a spaxel with a specific structure, but as a result fewer spaxels will be included in a structure, and so the patchier that structure will be. As an example of this we show in Figure \ref{fig:spiralthresholds} spiral masks as a range of thresholds for some example galaxies. This trade-off is particularly challenging for the spiral arm masks, but is present in all output from GZ:3D (and indeed in any crowd-sourced data, where there is always the risk of spurious classifications from volunteers who mis-click or misunderstand the task). 
 
  An alternative strategy is to perform your analysis once per volunteer mask (typically 15 times) with each galaxy in your sample, and then average the results afterwards. 
  
 In Table \ref{tab:threshold} we provide a summary of threshold choices made in all previous work (including this paper) which have used GZ:3D masks or clusters. We also comment on reasons for specific choices of the threshold. Note that number thresholds are not always identical to fraction thresholds as (see Section \ref{sec:methods}) while most masks are aggregated from 15 entries, there is some scatter in that total number. Making similar choices may aid in comparison with earlier work, however there is no single right answer to this choice; we recommend users inspect how their results change with differing thresholds, and indicate clearly the choice that is made. 
 
\begin{figure}
\includegraphics[angle=0,width=8cm]{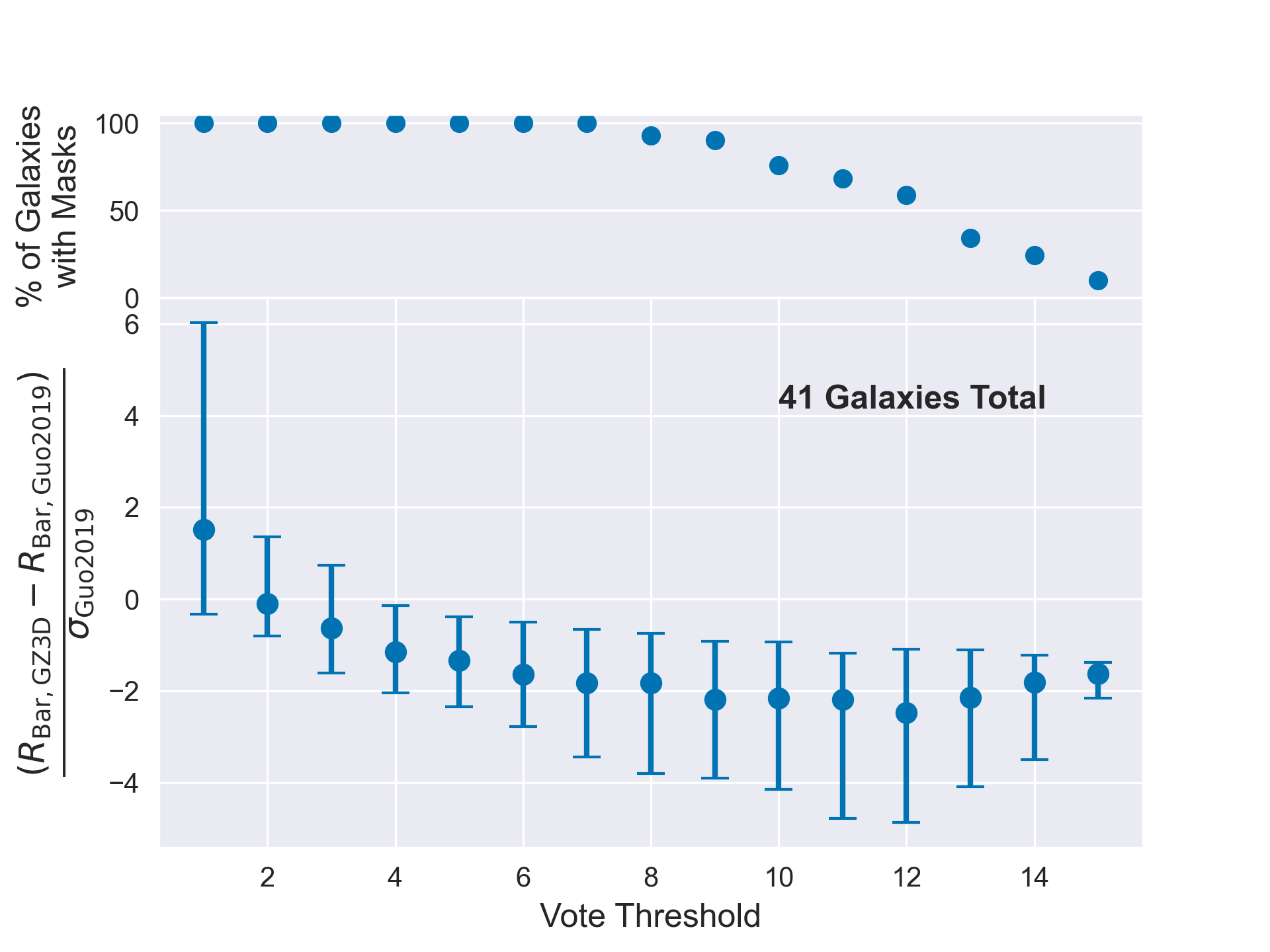}
\caption{Lower panel: A comparison of the bar lengths measured when different GZ:3D vote thresholds are used for a sample of 41 galaxies with bar lengths also measured by \citet{Guo2019} using a Fourier technique. The errors are 1$\sigma$. As the mask threshold increases, GZ:3D bar lengths decrease, settling at about 2$\sigma$ smaller than the Guo measurements once $N>6$ (or 40\%). Upper panel: the percent of all bar masks that have a mask with this threshold.}
\label{fig:barlength}
\end{figure}

 As an example of this process, we present in Figure \ref{fig:barlength} a comparison of the lengths of GZ:3D bars, indicated by fitting a minimum bounding box to masks at different thresholds, with bar lengths measured using a Fourier technique by \citet{Guo2019}. In total there are 41 galaxies with both measures. This, plot, which is similar to the analysis also presented in Appendix A of \citet{Krishnarao2020}, demonstrates how as the mask threshold increases, GZ:3D bar lengths decrease, however so does the number of galaxies with usable bar masks. We find that the bar length settles at a value about 2$\sigma$ lower than the Guo measurements once $N>6$ (or 40\%); for values which match Guo a threshold within 1$\sigma$ a lower threshold of $N=3$ (20\%) is recommended. 
 
\begin{table*}
\caption{\label{tab:threshold}Example thresholds (fraction of volunteers to identified a spaxel with a specific feature) used to create GZ:3D Masks}
\begin{tabular}{lcll}
Task & Threshold & Comment & Reference \\
\hline
Galaxy Centres & $N>6$ (40\%) & 85\% of all galaxy centres meet this & Section \ref{sec:centres}  \\
\hline 
Foreground Stars & $N>10$ (67\%) & Conservative limit; all locations have stellar-like spectra. & Section \ref{sec:stars} \\
& & Threshold used to flag stars in the DRP/DAP &  \citet{Westfall2019} \\
\hline
Bars & 20\% & Threshold in plot example.  & \citet{FraserMcKelvie2019}\\
 Bars & 80\%  & To be confident of bar ID. & \citet{FraserMcKelvie2019}\\
& 80\% & To be highly confident spaxel identified is in the bar. & \citet{FraserMcKelvie2020}\\
 & 20\% & Tested bar length with different thresholds & \citet{Krishnarao2020} \\
 & & This choice is compromise between accuracy and completeness. \\
 & 40\% & Informed by \citet{Krishnarao2020} for best length measurement.  & D. Krishnarao et al. in prep. \\
 \hline
Spirals & $N>3$ (20\%) & Creates continuous and distinct spirals & Section \ref{sec:spiralSF} \\
  & 25\% & To indicate representative region of arms & \citet{Peterken2019Density}\\
  & $<20$\% or $>40$\% & To indicate inter-arm and arm regions  & \citet{Peterken2019TimeSlice} \\
  & 50\% & Choice is compromise between accuracy and completeness & \citet{Greener2020} \\
  \hline 
\end{tabular}
\end{table*}

\subsection{Error on Binned Quantities}

As discussed extensively in \citet{Westfall2019} and \citet{Law2016}, the spatial binning scheme for MaNGA results in significant covariance between the values of adjacent 0.5\arcsec ~spaxels, since the spatial resolution of MaNGA is close to 2.5\arcsec, or 5 spaxels wide [also see \citet{Liu2020} for a discussion of regridding to account better for covariance]. When making use of data in bins provided by the DAP this issue is corrected for; however for users binning in GZ:3D's bins, set by the masks which identify certain features, more attention must be paid, as the errors provided per spaxel cannot be assumed to be independent and simply summed in quadrature. Instead the covariance needs to be accounted for, so that the relative error on a value constructed from the sum of $n$ spaxels is
\be
\sigma_{\rm tot}^2 = \sum_{i} \sigma_i^2 + \sum_i \sum_{(j\ne i)} \rho_{ij} \sigma_i \sigma_j,
\ee
where $\sigma_{i,j}$ are the relative error in a single spaxel, and $\rho_{ij}$ is the spatial covariance between two spaxels. 

 In principle, the level of spatial covariance between pixels depends on wavelength, but \citet{Westfall2019} find that the dependence on wavelength is rather weak, and provide an analytic fit for covariance as a function of the distance in spaxels between two locations $D_{ij}$ of
 \be
 \rho_{ij} = \exp^{-\frac{1}{2}\left( \frac{D_{ij}}{1.92}\right)^2},
 \ee
 {\it i.e.} a Gaussian of width 1.92 spaxels. For practical implementation it is found that setting this to $\rho_{ij}=0$ for $D_{ij}>6.4$ spaxels is reasonable. Code to implement these corrections is provided in the MaNGA DAP \citep{Westfall2019}\footnote{\url{https://sdss-mangadap.readthedocs.io/en/latest/spatialcovariance.html}}, we reiterate this subtlety here to ensure users of GZ:3D binning schemes are aware of the need to account for it. 

\section{Summary and Conclusions} \label{sec:summary}

In this paper, we have presented the {\it Galaxy Zoo: 3D} (GZ:3D) project, which has crowd-sourced the identification of galaxy centres and foreground stars, as well as the creation of spaxel masks which indicate the location of bars and spiral arm structures in MaNGA target galaxies. 

 We describe the building of the citizen science project on the {\it Zooniverse} platform, the SDSS $gri$ images of galaxies which were used, and how volunteer input was converted into clustered galaxy centre and foreground star points (with error regions), as well as pixel count masks for the location of bars and spirals. 
 
  We demonstrate how the galaxy centre markings can be used to find interesting MaNGA targets, demonstrating this with two bundles found with large numbers of galaxies. 
 
  The identification of foreground stars by GZ:3D has been used as an input to the MaNGA Data Analysis Pipeline to flag locations of likely foreground stars. For this application, a conservative cut was made, with only stars identified by at least $N=10$ volunteers being flagged. This was a deliberate choice to avoid accidentally flagging point-like components of target galaxies (e.g. HII regions).
 
 We provide a summary of published MaNGA results which have used the bar and/or spiral arm masks \citep[][D. Krishnarao in prep.]{Peterken2019Density,Peterken2019TimeSlice,FraserMcKelvie2019,FraserMcKelvie2020,Greener2020,Krishnarao2020}, as well as a guide to aid the reader in considerations needed when making use of GZ:3D crowd-sourced spaxel masks. As an illustration, we make use of the spiral masks to measure that on average star formation is enhanced by a factor of 3 in spiral arms relative to inter-arm regions,  as well as revealing how this fraction depends on radius in a sample of 825 spiral galaxies.

 The GZ:3D technique was developed to leverage human pattern recognition, and provide a guide to the location of complex internal structures in MaNGA data. GZ:3D makes it very easy to extract spectra from specific regions in the MaNGA data cubes and maps, however it doesn't take account of contamination from other components (i.e. the spectra from any given spaxel will in reality include light from multiple components). While spectroscopic structural decomposition (e.g. \citealt{Tabor2019}) is excellent for this, we do not the have the ability to apply it to any structures other than bulges and discs and spirals are challenging to fit even in more simple photometric decompositions (as discussed in \citealt{Lingard2020}, who test a crowd-sourced solution to this problem). 
  
 Spiral arms are particularly challenging to identify. Automated or machine learning methods to identify the locations of spiral arms are in the early stages of development \citep{Bekki2021} and struggle with anything other than $N=2$ bi-symmetric smooth arms. The masks provided here for a large sample of spiral galaxies may provide a useful training set for such efforts. 
 
 While data exploration and visualisation of multi-dimensional data, such as that provided in IFU surveys like MANGA, is complex, it is also a rich source to understand galaxies in our Universe. In this article, we have demonstrated how the technique of crowd-sourcing can be leveraged to help with the identification of regions of interest in astronomical images. 
   
\section*{Data Availability}
 All final GZ:3D results, which cover all MaNGA target galaxies, will be published as a Value Added Catalog (VAC) in the Seventeenth Release of the SDSS (DR17; SDSS Collaboration in prep.)\footnote{DR17 VACs are listed at \url{https://www.sdss.org/dr17/data_access/value-added-catalogs/}}, and made available to use via {\it Marvin} \citep{Cherinka2019}. For more details on how to use these data, see Section \ref{sec:data} above. Any scientists interested in results from the first phase of GZ:3D (which we do not recommend be used for science) should contact the first author.


\paragraph*{ACKNOWLEDGEMENTS.} 
This publication uses data generated via the \url{Zooniverse.org} platform, development of which is funded by generous support, including a Global Impact Award from Google, and by a grant from the Alfred P. Sloan Foundation. This publication has been made possible by the participation of almost 6000 volunteers in the {\it Galaxy Zoo: 3D} project on \url{Zooniverse.org.}

Funding for the Sloan Digital Sky Survey IV has been provided by the Alfred P. Sloan Foundation, the U.S. Department of Energy Office of Science, and the Participating Institutions. SDSS acknowledges support and resources from the centre for High-Performance Computing at the University of Utah. The SDSS web site is \url{www.sdss.org}.

SDSS is managed by the Astrophysical Research Consortium for the Participating Institutions of the SDSS Collaboration including the Brazilian Participation Group, the Carnegie Institution for Science, Carnegie Mellon University, centre for Astrophysics | Harvard \& Smithsonian (CfA), the Chilean Participation Group, the French Participation Group, Instituto de Astrof\'isica de Canarias, The Johns Hopkins University, Kavli Institute for the Physics and Mathematics of the Universe (IPMU) / University of Tokyo, the Korean Participation Group, Lawrence Berkeley National Laboratory, Leibniz Institut f\"ur Astrophysik Potsdam (AIP), Max-Planck-Institut f\"ur Astronomie (MPIA Heidelberg), Max-Planck-Institut f\"ur Astrophysik (MPA Garching), Max-Planck-Institut f\"ur Extraterrestrische Physik (MPE), National Astronomical Observatories of China, New Mexico State University, New York University, University of Notre Dame, Observat\'orio Nacional / MCTI, The Ohio State University, Pennsylvania State University, Shanghai Astronomical Observatory, United Kingdom Participation Group, Universidad Nacional Aut\'onoma de M\'exico, University of Arizona, University of Colorado Boulder, University of Oxford, University of Portsmouth, University of Utah, University of Virginia, University of Washington, University of Wisconsin, Vanderbilt University, and Yale University.

We gratefully acknowledge the National Science Foundation's support of the Keck Northeast Astronomy Consortium's REU program through grants AST-1005024 and AST-1950797, the KINSC (Koshland Integrated Natural Sciences centre) at Haverford College for Summer Scholar funding, and the Ogden Trust, UK for support for summer undergraduate internships. 

The work presented in Section 3.5, and Figures 7, 8 and~9 was part of the Haverford College Senior Thesis of Shoaib Shamsi '21.


\bibliography{references}{}
\bibliographystyle{mnras}
\label{lastpage}  
\end{document}